\documentclass[aps,prd,reprint,twocolumn,superscriptaddress,preprintnumbers,nofootinbib]{revtex4-1}

\usepackage{amsthm}
\usepackage{amsmath}
\usepackage{graphicx}
\usepackage{slashed}
\usepackage{amssymb}
\usepackage{dsfont}
\usepackage{float}
\usepackage[colorlinks=True, citecolor=blue, urlcolor=blue, linkcolor=blue]{hyperref}
\usepackage[braket,qm]{qcircuit}
\usepackage{lineno}
\usepackage[dvipsnames]{xcolor}

\usepackage{placeins}

\newcommand\pTmiss{\ensuremath{p_T^\mathrm{miss}}}

\newcommand\GeVsquare{\ensuremath{\mathrm{GeV}^{2}}}

\newcommand\nn{\nonumber}

\def\bea#1\eea{\begin{align}#1\end{align}} 
\def\bea#1\eea{\begin{align}#1\end{align}}
\usepackage{slashed}
\newcommand{\nnu}{\nonumber\\}

\newcommand\diff{{\rm d}}
\begin{document}
\title{Neutrino-tagged jets at the Electron-Ion Collider}
\author{Miguel Arratia}
\affiliation{Department of Physics and Astronomy, University of California, Riverside, CA 92521, USA}
\affiliation{Thomas Jefferson National Accelerator Facility, Newport News, VA 23606, USA}

\author{Zhong-Bo Kang}
\affiliation{Department of Physics and Astronomy, University of California, Los Angeles, CA 90095, USA}
\affiliation{Mani L. Bhaumik Institute for Theoretical Physics, University of California, Los Angeles, CA 90095, USA}
\affiliation{Center for Frontiers in Nuclear Science, Stony Brook University, Stony Brook, NY 11794, USA}

\author{Sebouh J. Paul}
\affiliation{Department of Physics and Astronomy, University of California, Riverside, CA 92521, USA}

\author{Alexei Prokudin}
\affiliation{Division of Science, Penn State University Berks, Reading, PA 19610, USA}
\affiliation{Thomas Jefferson National Accelerator Facility, Newport News, VA 23606, USA}

\author{Felix Ringer}
\affiliation{Thomas Jefferson National Accelerator Facility, Newport News, VA 23606, USA}
\affiliation{Department of Physics, Old Dominion University, Norfolk, VA 23529, USA}
\affiliation{C.N. Yang Institute for Theoretical Physics, Stony Brook University, Stony Brook, NY 11794,USA}
\affiliation{Department of Physics and Astronomy, Stony Brook University, Stony Brook, NY 11794, USA}

\author{Fanyi Zhao}
\affiliation{Department of Physics and Astronomy, University of California, Los Angeles, CA 90095, USA}
\affiliation{Mani L. Bhaumik Institute for Theoretical Physics, University of California, Los Angeles, CA 90095, USA}
\affiliation{Center for Frontiers in Nuclear Science, Stony Brook University, Stony Brook, NY 11794, USA}

\date{\today} 
\preprint{JLAB-THY-22-3758, YITP-SB-2022-38}

\begin{abstract}
We explore the potential of jet observables in charged-current deep inelastic scattering (CC DIS) events at the future Electron-Ion Collider (EIC). Tagging jets with a recoiling neutrino, which can be identified by the event's missing transverse momentum, will allow for flavor-sensitive measurements of Transverse Momentum Dependent parton distribution functions (TMDs). We present the first predictions for transverse-spin asymmetries in azimuthal neutrino-jet correlations and hadron-in-jet measurements. We study the kinematic reach and the precision of these measurements and explore their feasibility using parameterized detector simulations. We conclude that jet production in CC DIS, while challenging in terms of luminosity requirements, will complement the EIC experimental program to study the three-dimensional structure of the nucleon encoded in TMDs. 
\end{abstract}

\maketitle

\section{Introduction}
\label{sec:outline}

The Electron-Ion Collider (EIC) will usher in a new era for the study of the 3D structure of the nucleon~\cite{Accardi:2012qut,AbdulKhalek:2021gbh}. Its high luminosity and polarization of both electron and hadron beams will enable precise measurements of observables related to Transverse Momentum Dependent parton distribution and fragmentation functions (TMDs). 

Jets are energetic sprays of particles observed at high-energy collider experiments that are closely related to the underlying quark and gluon dynamics of hard-scattering events. Jets at the EIC will have transverse momenta up to $\sim 40$~GeV~\cite{Arratia:2019vju,Page:2019gbf}. The EIC will produce the first jets in deep-inelastic scattering off transversely-polarized nucleons. The potential of jets produced in neutral-current deep-inelastic scattering (NC DIS) has recently been explored, see \textit{e.g.}~Refs.~\cite{Hinderer:2015hra,Boughezal:2018azh,Zheng:2018ssm,Gutierrez-Reyes:2018qez,Gutierrez-Reyes:2019vbx, Liu:2018trl,Kang:2020xyq,Borsa:2020yxh,Arratia:2020azl,Arratia:2020nxw,Liu:2020dct,Arratia:2020ssx,Kang:2020fka,Guzey:2020gkk,Kang:2021ffh,delCastillo:2021znl,Makris:2021drz,Zhang:2021tcc,AbdulKhalek:2022hcn,Tong:2022zwp,Liu:2022wop,Lee:2022kdn}. In this work, we will focus on jets produced in charged-current deep-inelastic scattering (CC DIS).

\begin{figure}
    \centering
    \vspace*{-1.3cm}    \includegraphics[width=0.75\columnwidth]{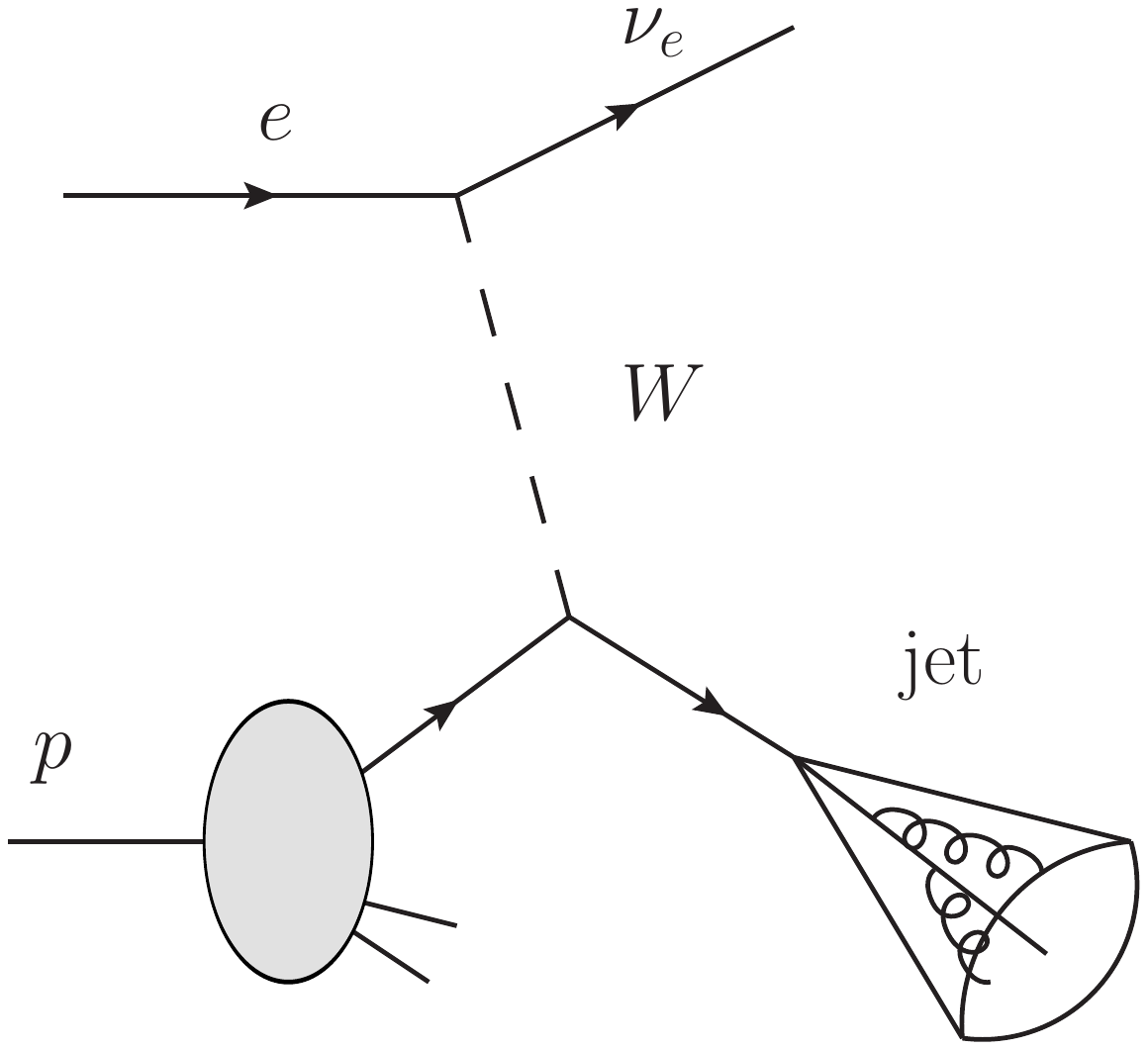}
    \vspace*{-1.3cm}
\caption{Charged-current deep-inelastic scattering where the produced jet recoils against a neutrino.\label{fig:diagram}}
\end{figure}

The CC DIS process, which involves the exchange of a virtual $W^\pm$ boson, enables jet measurements that are sensitive to the flavor of the scattered quark. The leading-order process, $W^\pm q\to q'$, is illustrated in Fig.~\ref{fig:diagram}. 
Due to the conservation of electric charge, electrons can only scatter via the exchange of a $W^-$ off positively charged partons, which are predominantly $u$-quarks, especially at large $x$.  Likewise, with a positron beam, scattering processes occur predominantly with $d$-quarks through the exchange of a $W^+$ boson. Moreover, tagging either charm or strange jets can further enhance the flavor sensitivity of jet measurements~\cite{Zheng:2018ssm,Arratia:2020azl,Kang:2020fka,Arrington:2021yeb,Lee:2022kdn}.

The H1 and ZEUS collaborations measured inclusive CC DIS off unpolarized protons with longitudinally polarized electron and positron beams~\cite{Aaron:2012qi,Collaboration:2010xc,Chekanov:2008aa,Chekanov:2002zs,Breitweg:1999aa,Derrick:1996sw}. These measurements allowed for constraining the flavor dependence of collinear parton distribution functions (PDFs)~\cite{Abramowicz:2015mha}. In addition, CC DIS jet production measurements by the ZEUS collaboration~\cite{Chekanov:2003jd,Chekanov:2008af}, were compared to precise next-to-next-to-next-to-leading order QCD calculations~\cite{Gehrmann:2018odt}.

One of the main challenges in measuring CC DIS is the
measurement of the events' kinematic variables, Bjorken $x$ and $Q^{2}$, in the presence of an undetected final-state neutrino. Several methods exist to address this challenge~\cite{Blumlein:2012bf,Collaboration:2010xc}. The feasibility studies of CC DIS at the EIC have been performed in Refs.~\cite{Aschenauer:2013iia,AbdulKhalek:2021gbh} for DIS off longitudinally polarized protons with the goal to access helicity PDFs. In this study we will focus on the CC DIS off transversely polarized protons that will lead to measurements of the transverse-spin effects related to TMDs. 

In semi-inclusive DIS (SIDIS), transverse-spin asymmetries can be extracted from modulations of the azimuthal angles with respect to the virtual-boson direction, typically in the Breit frame~\cite{Avakian:2019drf,Anselmino:1993tc,Chen:2020ugq,PhysRevD.103.016011,Yang:2022sbz}. In CC DIS, this approach requires a measurement of the three-momentum of the scattered neutrino to define the azimuthal angle, which is challenging due to acceptance losses at forward rapidities.

Jet-based measurements of spin asymmetries can reduce these difficulties. Following Liu \textit{et al}.~\cite{Liu:2018trl}, TMDs can be accessed in lepton-jet azimuthal correlation measurements in the laboratory frame instead of the conventional Breit frame. Liu \textit{et al}.~\cite{Liu:2018trl} considered NC DIS, but the formalism can be extended to CC DIS as well. The advantage of this approach is that the measurement of the azimuthal angle only requires the neutrino's transverse momentum in the lab frame, which, in general, can be measured more precisely than the full three-momentum~\cite{Gao:2022bzi}. In addition, hadron-in-jet asymmetry measurements can be performed by defining an azimuthal angle of the hadron with respect to the jet axis~\cite{Yuan:2007nd,Procura:2009vm,Kang:2017glf,Arratia:2020nxw,Kang:2021ffh}.  

The jet-based TMD measurements have the additional advantage of decoupling initial- and final-state TMD effects (at leading power in the jet radius)~\cite{Kang:2021ffh}. That is, they do not involve a convolution of TMD PDFs and fragmentation functions which can introduce strong correlations in global fits of SIDIS data~\cite{Bacchetta:2017gcc,Scimemi:2019cmh}. 

In this paper, we present the first study of neutrino-jet and hadron-in-jet spin asymmetries in CC DIS. We determine all possible spin asymmetries within the TMD factorization formalism. We present numerical estimates for transverse single-spin asymmetries in CC DIS. We also perform feasibility studies of these measurements using fast detector simulations and we quantify the expected kinematic reach and the statistical uncertainties. 

The remainder of this paper is organized as follows.  We describe the proposed measurements in
Section~\ref{sec:proposedmeasurement} and the theoretical framework in Section~\ref{sec:theory}. We describe the fast detector simulation in Section~\ref{sec:simulation} and the expected experimental performance in Section~\ref{sec:resolution}.  We estimate the background in Section~\ref{sec:backgrounds}, and show projections for the transverse-spin asymmetries in Section~\ref{sec:stat}. We conclude in Section~\ref{sec:Conclusions}.

\section{Proposed measurements}
\label{sec:proposedmeasurement}

Following Liu \textit{et al}.~\cite{Liu:2018trl}, we propose the measurement of the distribution of the azimuthal separation between the outgoing neutrino (as determined from the missing transverse momentum), and the jet. Due to the momentum conservation, the jet and neutrino are expected to be predominantly produced back-to-back. Therefore, the azimuthal distribution is expected to be centered around $\phi_{\rm jet}-\phi_\nu-\pi=0$, with some finite width due to out-of-cone QCD radiation and the non-zero initial momentum of the scattered quark. In the next section, we determine all spin asymmetries that can be measured in neutrino-jet and hadron-in-jet production within TMD factorization.

Moreover, we propose to measure the transverse single-spin asymmetry in neutrino-jet correlations, also known as the left-right asymmetry
\begin{equation}
A_{\mathrm{UT}} = \frac{{\rm d}\sigma^{\uparrow} - {\rm d}\sigma^{\downarrow}}{{\rm d}\sigma^{\uparrow} + {\rm d}\sigma^{\downarrow}}\,,
\end{equation}
where, ${\rm d}\sigma^{\uparrow,\downarrow}$ refers to the differential cross section measured with transverse polarization of the initial proton pointing up or down. This asymmetry is expected to exhibit a modulation with respect to the angular separation between the incoming proton spin, $\phi_S$, and the momentum imbalance, $\phi_{q}$, \textit{i.e.},
\begin{equation}
     A_{\mathrm{UT}} = A^{\sin(\phi_S-\phi_q)}_{\mathrm{UT}} \sin (\phi_S-\phi_q).
     \label{eq:AUT}
\end{equation}
Here, the momentum imbalance between the jet and the neutrino is defined by $\vec q_T = \vec p_T^{\,\rm jet}+\vec p_T^{\,\nu}$. This asymmetry is sensitive to the Sivers function~\cite{Liu:2018trl,Liu:2020dct}, which describes the anisotropy of unpolarized partons in a transversely polarized proton. 

We also propose to perform a hadron-in-jet measurement in CC DIS of the asymmetry $A^{\sin (\phi_S-\phi_{\rm h})}_{\mathrm{UT}}$  defined for the azimuthal angle of the hadron in jet $\phi_{\rm h}$. In NC DIS, the hadron-in-jet asymmetry is sensitive to both the Collins and the transversity functions~\cite{Arratia:2020nxw}.

\section{Theoretical framework}
\label{sec:theory}

In this section, we discuss inclusive jet production in CC DIS, neutrino-jet correlations, and hadron-in-jet observables.

\subsection{Inclusive jet production}

We follow the theoretical framework developed in Refs.~\cite{Liu:2018trl,Arratia:2020nxw,Liu:2020dct,Kang:2021ffh} for the NC DIS. At the parton level, we consider the leading-order process $eq \rightarrow \nu q'$ mediated via the exchange of a virtual $W$ boson. We consider the cross section differential in Bjorken $x$ and the transverse momentum of the produced neutrino, $p_T^\nu$, which is defined relative to the beam direction in the laboratory frame. The leading-order cross section can be written as
\bea\label{eq:crosssection0}
    &\frac{\diff \sigma^{ep\rightarrow \nu\mathrm{jet}X}}{\diff x\,\diff^2 \vec p^{\; \nu}_T} =  \,\sum_q \sigma_0^{eq\rightarrow \nu q'} f_q(x,\mu)\,,
\eea
where the renormalization scale $\mu$ of the PDF $f_q$ is chosen at the order of the hard scale of the process $\mu\sim p_T^\nu$. 
The prefactor $\sigma_0$ for initial quarks $u$ and $\bar{d}$ are given by
\bea
& \sigma_0^{eu\rightarrow\nu d}=\frac{|\overline{\mathcal{M}}_{eu\rightarrow\nu d}|^2}{16\pi^2\hat{s}^2}\frac{\hat{t}}{x(\hat{t}-\hat{u})}\nnu
&=8(G_F m_W^2)^2|V_{ud}|^2\frac{\hat{s}^2}{(\hat{t}-m_W^2)^2+m_W^2\Gamma_W^2}\frac{\hat{t}}{x(\hat{t}-\hat{u})}\,,\label{eq:sigma1}\\
&\sigma_0^{e\bar{d}\rightarrow\nu \bar{u}}=\frac{|\overline{\mathcal{M}}_{e\bar{d}\rightarrow\nu \bar{u}}|^2}{16\pi^2\hat{s}^2}\frac{\hat{t}}{x(\hat{t}-\hat{u})}\nnu
&=8(G_F m_W^2)^2|V_{ud}|^2\frac{\hat{u}^2}{(\hat{t}-m_W^2)^2+m_W^2\Gamma_W^2}\frac{\hat{t}}{x(\hat{t}-\hat{u})}\,,\label{eq:sigma2}
\eea
where $G_F$ is the Fermi constant, $m_W$ and $\Gamma_W$ are the $W$ boson mass and decay width, and $V_{ud}$ is the standard CKM matrix element. Here ${\hat{t}}/(x(\hat{t}-\hat{u}))$ is the Jacobian factor, which is obtained by transforming the cross section to be differential in Bjorken-$x$ instead of the neutrino rapidity $y_\nu$. These two variables are related by
\bea
x=\frac{p_T^\nu e^{y_\nu}}{\sqrt{s}-p_T^\nu e^{-y_\nu}} .
\eea
The Mandelstam variables in Eqs.~\eqref{eq:sigma1} and~\eqref{eq:sigma2} can be written in terms of the kinematic variables of the produced neutrino and the center-of-mass energy, namely
\bea
&\hat{s}=x s \,,\\
&\hat{t}=-Q^2=-\sqrt{s} p_T^\nu e^{y_\nu}=-x \sqrt{s} p_T^{\mathrm{jet}} e^{-y_{\mathrm{jet}}}, \\
&\hat{u}=-x \sqrt{s} p_T^\nu e^{-y_\nu}=-\sqrt{s} p_T^{\mathrm{jet}} e^{y_{\mathrm{jet}}}\,.
\eea
Here $p_T^{\text {jet }}$ and $y_{\text {jet }}$ denote the jet transverse momentum and rapidity, respectively.

\subsection{Neutrino-jet correlations}

Next, we discuss neutrino-jet correlations via the exchange of a $W^-$ boson  in polarized electron-proton scattering
\bea
p(P_A,\lambda_p,\vec{S}_T)+e(P_B,\lambda_e)\rightarrow \mathrm{jet}(P_J)+\nu(P_D)+X \,.
\eea
Here $\lambda$ indicates the longitudinal polarization and $\vec S_T$ denotes the transverse spin vector of the proton. In order to access TMDs, we study back-to-back neutrino-jet production in the $ep$ collision frame. By defining light-cone vectors $n_+^\mu=\frac{1}{\sqrt{2}}(1,0,0,1)$ and $n_-^\mu=\frac{1}{\sqrt{2}}(1,0,0,-1)$, we write the momentum of the incoming proton $P_A$ and the electron $P_B$ as
\bea
P_A^\mu&=P^+n_+^\mu+\frac{M^2}{2P^+}n_-^\mu\approx P^+n_+^\mu\ = \sqrt{\frac{s}{2}}n_{+}^\mu\, ,\\
P_B^\mu&=\sqrt{\frac{s}{2}}n_{-}^\mu\, .
\eea
Here $s=(P_A+P_B)^2$ is the center-of-mass energy. We set the final observed jet to be produced in the $xz$-plane, with the following momentum $P_J^\mu = E_J\left(1, \sin\theta_J, 0, \cos\theta_J\right)$. Here $E_J$ is the jet energy and the angle $\theta_J$ is measured with respect to the beam direction. 
We find that the differential cross section can be written in terms of the structure functions as follows~\footnote{Notice that unlike the usual practice for SIDIS or DY cross sections, we do not factor out the elementary cross-sections from the structure functions. Our structure functions therefore become dimension-full quantities, see for instance Ref.~\cite{Kang:2022dpx}.}
\bea
&\frac{{\rm d}\sigma^{ep\rightarrow \nu\mathrm{jet}X}}{{\rm d}x\, {\rm d}^2\vec{p}^{\,\nu}_{T}\, {\rm d}^2q_T}=F_{UU}+\lambda_p F_{UL}\nnu
&\hspace{1cm}+|S_T|\left[\sin(\phi_{q} -\phi_{S_A})F_{UT}^{\sin(\phi_{q}-\phi_{S_A})}\right.\nnu
&\hspace{2.2cm}\left.+\cos(\phi_{q}-\phi_{S_A})F_{UT}^{\cos(\phi_{q}-\phi_{S_A})}\right]\nnu
&\hspace{1cm}+\lambda_e\left[F_{LU}+\lambda_pF_{LL}\right.\nnu
&\hspace{1.9cm}+|S_T|\sin(\phi_{q}-\phi_{S_A})F_{LT}^{\sin(\phi_{q}-\phi_{S_A})}\nnu
&\hspace{1.9cm}\left.+|S_T|\cos(\phi_{q}-\phi_{S_A})F_{LT}^{\cos(\phi_{q}-\phi_{S_A})}\right]\,.\label{eq:crosssection1}
\eea
The subscripts $E$ and $P$ of a structure function $F_{EP}$ indicate the polarization of the incoming electron and incoming proton respectively: $U$ for unpolarized, $L$ for longitudinally polarized or $T$ for transversely polarized. For example, for unpolarized scattering, the structure function is denoted by $F_{UU}$. In the limit of small values of the transverse-momentum imbalance $|\vec q_T|\ll p_T^{\rm jet}\sim p_T^\nu$, one can write this structure function in the following form using the TMD factorization formalism
\bea
F_{UU}=&\sum_q \frac{|\overline{\mathcal{M}}_{eq\rightarrow\nu q'}|^2}{16\pi^2\hat{s}^2}H(Q,\mu)\,\mathcal{J}_q(p_T^{\mathrm{jet}}R,\mu)\nnu
&\times\int\frac{{\rm d} b_T b_T}{2\pi}J_0(q_Tb_T)f_1^{\mathrm{TMD}}(x,b_T,\mu, \zeta)\nnu
& \qquad\times S_q(b_T,y_{\mathrm{jet}},R,\mu)\, .
\label{eq:factorization1}
\eea
Here $H(Q,\mu)$ is the hard function, which accounts for virtual corrections at the hard scale $Q$. The jet function $\mathcal{J}_q$ is associated with collinear dynamics of the jet with characteristic scale $\mu_J\sim p_T^{\rm jet}R$~\cite{Ellis:2010rwa}. For our numerical results presented below we use the anti-$k_T$ algorithm~\cite{Cacciari:2008gp}. The quark TMD PDF including the appropriate soft factor for a generic TMD in $b_T$-space is defined by~\cite{Boer:2011xd,Kang:2021ffh}
\bea\label{eq:ftb}
{f}_q^{(n),{\rm TMD}}(x,b_T,\mu, \zeta)=&\frac{2 \pi n !}{\left({M^{2}}\right)^{n}} \int {\rm d}k_T \, k_T\left(\frac{k_T}{b_T}\right)^{n}J_{n}\left(k_T b\right) \nnu
&\times  \tilde{f}_q^{\,\rm TMD}\!\left(x, k_T^2,\mu, \zeta \right)\,,
\eea
where $M$ is the mass of the nucleon and $J_n$ is the $n$-th order Bessel function. Here $\mu$ is the renormalization scale, while $\zeta$ is the so-called Collins-Soper scale~\cite{Collins:2011zzd}. Notice that for the unpolarized TMD $f_1^{\rm TMD}$, $n=0$. The remaining soft function $S_{q}$ in Eq.~(\ref{eq:factorization1}) includes a contribution from the global soft function which depends on the Wilson lines in the beam and jet directions, and the collinear-soft function associated with the soft jet dynamics. Since $S_q$ accounts for different soft contributions, it depends on both the jet rapidity $y_{\rm jet}$ and the jet radius $R$ and the expression of $S_q$ is given in~\cite{Arratia:2020nxw}.

The other cross sections or structure functions in Eq.~(\ref{eq:crosssection1}) depend on the polarization state of the nucleon  (longitudinal or transverse) and the longitudinal lepton polarization. As an example, we consider the case where the initial proton is transversely polarized. The cross section or structure function $F_{UT}^{\sin(\phi_{q}-\phi_{S_A})}$ sensitive to correlations of an unpolarized quark in the transversely polarized proton is given by
\bea
F_{UT}^{\sin(\phi_{q}-\phi_{S_A})}=&\sum_q \frac{|\overline{\mathcal{M}}_{eq\rightarrow\nu q'}|^2}{16\pi^2\hat{s}^2}H(Q,\mu)\,\mathcal{J}_q(p_T^{\mathrm{jet}}R,\mu)\nnu
&\times\int\frac{{\rm d}b_Tb_T^2}{4\pi M}J_1(q_Tb_T)f_{1T}^{\perp(1), \mathrm{TMD}}(x,b_T,\mu, \zeta)\nnu
& \qquad\times S_q(b_T,y_{\mathrm{jet}},R,\mu)\, .
\eea
In this case, the structure function is related to the Bessel function of the 1st order $J_1(q_Tb_T)$ and the Sivers function $f_{1T}^{\perp(1), \mathrm{TMD}}(x,b_T,\mu, \zeta)$ in $b_T$-space as defined in Eq.~\eqref{eq:ftb}.

We provide more details about the other structure functions in Eq.~(\ref{eq:crosssection1}) in the Appendix~\ref{app:jet}. For example, for an incoming electron with helicity $\lambda_e$, one replaces $\sigma_0^{eq\rightarrow\nu q'}$ with $\sigma_0^{e_Lq\rightarrow\nu q'}$ as given in Eq.~\eqref{eq:flu}. 

\subsection{Hadron distributions inside jets}

\begin{figure}
    \centering
    \includegraphics[width=0.90\columnwidth]{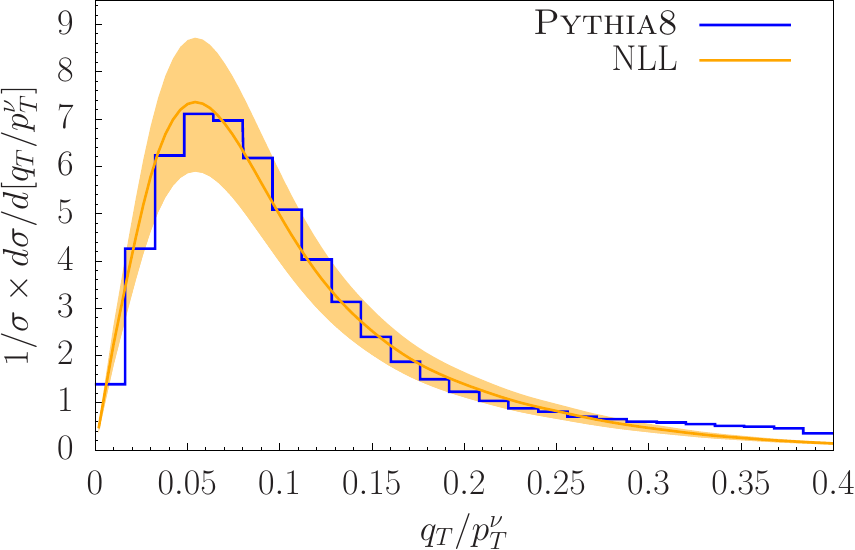}
\caption{Normalized distribution of the neutrino-jet imbalance momentum $q_T/p^\nu_T$ in unpolarized electron-proton scattering via the exchange of a $W^-$ boson.  We show our theoretical results at NLL accuracy with QCD scale uncertainties (orange) compared to Monte-Carlo event-generator simulations obtained with \textsc{Pythia}8~\cite{Sjostrand:2007gs} (blue).~\label{fig:qt_theory}}
\end{figure}

\begin{figure*}
    \centering
 \includegraphics[width=0.4\textwidth]{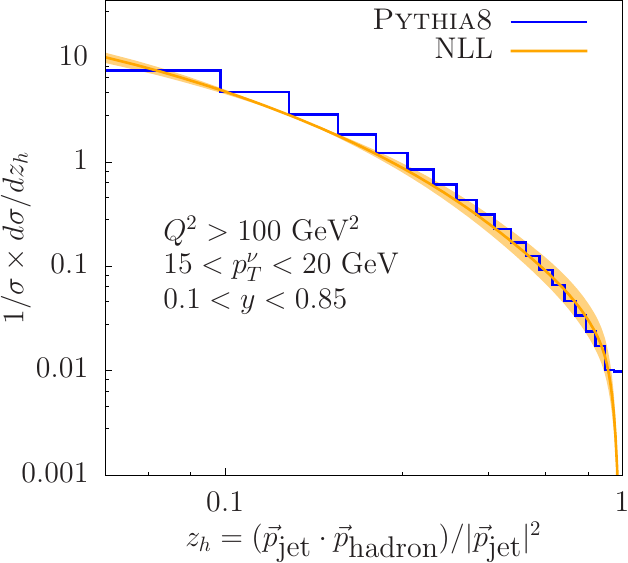}\hspace{0.05\textwidth}
  \includegraphics[width=0.4\textwidth]{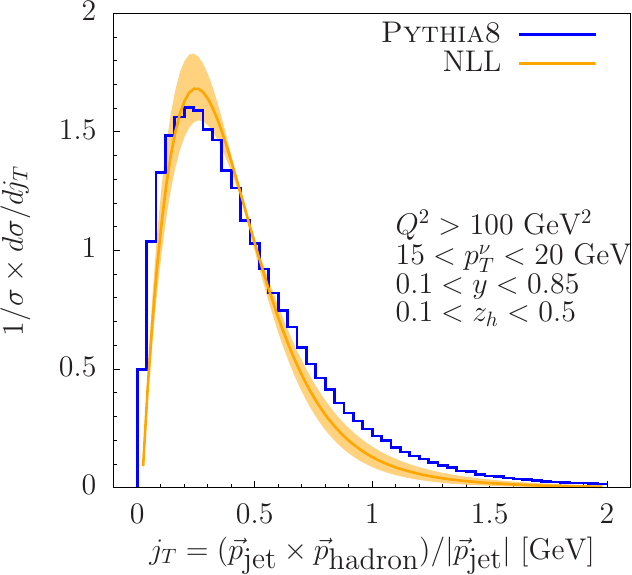}
\caption{Distributions of the $\pi^\pm$-in-jet longitudinal-momentum fraction $z_h$ (left) and the transverse momentum $j_T$ (right) in CC DIS events. We show our theoretical results at NLL accuracy with QCD scale uncertainties (orange) compared to Monte-Carlo event-generator simulations obtained with \textsc{Pythia}8~\cite{Sjostrand:2007gs} (blue).~\label{fig:zh_and_jt}}
\end{figure*}

In this subsection, we study the longitudinal- and transverse-momentum distributions of hadrons in the identified jet for electron-proton scattering via the exchange of a $W^-$ boson:
\bea
&p(P_A,\lambda_p,\vec{S}_T)+e(P_B,\lambda_e)\nnu
&\hspace{1.2cm}\rightarrow \big(\mathrm{jet}(P_J)h(z_h,\vec{j}_T)\big)+\nu(P_D)+X.
\eea

The production of unpolarized final-state hadrons at leading twist is encoded in two TMD jet fragmentation functions (JFFs), $\mathcal{D}_1$ and $\mathcal{H}_1^\perp$~\cite{Kang:2021ffh}
\bea
\Delta(z_h,\vec{j}_T)=\,&\mathcal{D}_1^{h/q}(z_h,{j}_T)\frac{\slashed{n}_-}{2}\nnu
&-i\mathcal{H}^{\perp,h/q}_1(z_h,{j}_T)\frac{\slashed{j_T}}{z_hM_h}\frac{\slashed{n}_-}{2}\,.
\eea
Thus, we find the following  differential cross section expressed in terms of structure functions
\bea
&\frac{{\rm d}\sigma^{ep\rightarrow \nu+\mathrm{jet}X}}{{\rm d}x\, {\rm d}^2\vec{p}^{\,\nu}_{T}\, {\rm d}^2q_T\, {\rm d}z_h\, {\rm d}^2{j_T}}=F^h_{UU}+\lambda_p F^h_{UL}\nnu
&\qquad+|S_T|\left[\cos(\phi_{q}-\phi_{S_A})F_{UT}^{h,\cos(\phi_{q}-\phi_{S_A})}\right.\nnu
&\quad\hspace{1.35cm}\left.+\sin(\phi_{q} -\phi_{S_A})F_{UT}^{h,\sin(\phi_{q}-\phi_{S_A})}\right]\nnu
&\qquad+\lambda_e\left[F^h_{LU}+\lambda_pF^h_{LL}\right.\nnu
&\qquad\hspace{1.1cm}+|S_T|\sin(\phi_{q}-\phi_{S_A})F_{LT}^{h,\sin(\phi_{q}-\phi_{S_A})}\nnu
&\qquad\hspace{1.1cm}\left.+|S_T|\cos(\phi_{q}-\phi_{S_A})F_{LT}^{\cos(\phi_{q}-\phi_{S_A})}\right]\,.\label{eq:diff_xs_hadron}
\eea

In total, we find 8 structure functions and the full expression is provided in Appendix~\ref{app:h-in-jet}. Following the same convention we used in Eq.~\eqref{eq:crosssection1}, the subscripts $E$ and $P$ of a structure function $F^h_{EP}$ here also indicate the polarization of the incoming electron and the incoming proton, respectively. Notably, we found that none of the Collins-type jet fragmentation functions contribute in Eq.~\eqref{eq:diff_xs_hadron}. The reason is that chiral-odd functions have to be coupled with another chiral-odd function, the chirality between two factors of $(1-\gamma_5)$ resulting from the weak charged-current vertices must be odd. As a result, see Eq.~(\ref{eq:g5mat}), we always have  $(1-\gamma_5)(1+\gamma_5)=0$, which implies that all terms involving chiral-odd functions vanish. This conclusion is robust at leading power of the TMD factorization formalism we are using. However, whether this still holds going beyond the leading power TMD factorization, see \textit{e.g.} within TMD factorization at sub-leading power~\cite{Ebert:2021jhy,Rodini:2022wki,Gamberg:2022lju} or including higher loops as discussed in Refs.~\cite{Benic:2019zvg,Benic:2021gya}, needs further investigation. We leave such a study to a future publication. 

Using TMD factorization at leading power, see Refs.~\cite{Yuan:2007nd,Procura:2009vm,Ellis:2010rwa,Jain:2011xz,Kaufmann:2015hma,Kang:2016ehg,Bain:2016rrv,Kang:2017glf,Kang:2017btw,Kang:2019ahe}, we can write the unpolarized structure function where a hadron is measured inside the jet as follows:
\begin{align}\label{eq:crosssection2}
  F^h_{UU}  &=\,
    H(Q,\mu) \sum_q\sigma_0^{eq\rightarrow\nu q'} \, {\cal D}_1^{h/q}(z_h, j_T,p_T^{\rm jet}R,\mu)
    \\ &\times \,
    \int\frac{\diff^2 \vec b_T}{(2\pi)^2}\, e^{i\vec q_T\cdot \vec b_T}\, f_1^{\rm TMD}(x, b_T,\mu, \zeta)\, S_{q}( b_T,y_{\rm jet},R,\mu) \,.\nn
\end{align}
Here the variables $z_h = \vec{p}_{h} \cdot \vec{p}_{\rm jet}/|\vec{p}_{\rm jet}|^{2}$ and ${j}_T=\left|\vec{p}_{h} \times \vec{p}_{\rm jet}\right|/|\vec{p}_{\rm jet}|^{2}$ denote the longitudinal momentum fraction and the transverse momentum relative to the (standard) jet axis of the hadron inside the jet, respectively. In the factorized cross section in Eq.~(\ref{eq:crosssection2}), ${\cal D}_1^{h/q}$ is a TMD fragmenting jet function. It describes the hadron-in-jet measurement and replaces the jet function $\mathcal{J}_q$ in Eq.~(\ref{eq:factorization1}). At next-to-leading logarithmic (NLL) accuracy, we can write ${\cal D}_1^{h/q}$ as
\bea
    &{\cal D}_1^{h/q}(z_h, j_T,p_T^{\rm jet}R,\mu)\nnu
    &\hspace{1.cm}=\int\frac{\diff^2\vec b_T^{\,\prime}}{(2\pi)^2}e^{{i\vec j_T\cdot\vec b_T^{\,\prime}}/{z_h}} D_1^{q/h}(z_h,\vec b_T^{\,\prime},p_T^{\rm jet}R) \,.\label{eq:TMDFF}
\eea
Here we work in Fourier conjugate space and $D_1^{q/h}$ is a TMD fragmentation function (TMDFF) evaluated at the jet scale. We use the Fourier variable $\vec b_T^{\,\prime}$ here to indicate that there is no convolution of the TMD fragmentation function with the TMD PDF in Eq.~(\ref{eq:crosssection2}). Also note that the TMDFFs can be matched to the collinear FFs~\cite{Aybat:2011zv,Collins:2011zzd} and in this work, we apply the extraction of collinear fragmentation functions in ~\cite{deFlorian:2014xna} for constructing the TMDFFs. See Ref.~\cite{Kang:2017btw} for more details. 
\section{Simulation}
\label{sec:simulation}

In this section, we discuss Monte-Carlo event-generator results for neutrino-jet correlations as well as detector-response simulations. We show comparisons between theoretical calculations discussed in the previous section and the Monte-Carlo simulations for unpolarized cross sections in CC DIS events.

\subsection{Event-Generation with \textsc{Pythia8}}
\label{sec:pythia}

We used \textsc{Pythia8}~\cite{Sjostrand:2007gs} to simulate CC DIS events in unpolarized electron-proton and positron-proton collisions.  
We choose the energies of the incoming electron and proton as 10~GeV and 275~GeV, respectively. These beam-energy values, which yield a center-of-mass energy of $\sqrt{s}=105$~GeV, correspond to the operation point that maximizes the luminosity of the EIC design~\cite{EICdesign}. Following Ref.~\cite{Aschenauer:2013iia}, we selected events with $Q^{2} > 100$ \GeVsquare. QED radiative effects~\cite{Badelek:1994uq,Liu:2021jfp} are not included in the simulation to match the calculations in Sec.~\ref{sec:theory}~\footnote{Based on similar measurements in NC DIS~\cite{H1:2021wkz}, the QED corrections are expected to be small for the observables considered in this work. Therefore, we do not expect that our conclusions are affected by these effects.}. We used the \textsc{Fastjet}3.3 package~\cite{Cacciari:2011ma} to reconstruct jets with the anti-$k_T$ algorithm~\cite{Cacciari:2008gp} and jet radius parameter $R=1$. The input particles for the generator-level jet-finding algorithm are all stable particles ($c\tau>$ 10~mm), except for  neutrinos.    

Figure~\ref{fig:qt_theory} shows our theoretical results at NLL accuracy for the transverse momentum imbalance of the neutrino and jet $q_T/p^\nu_T$. In addition, we show the \textsc{Pythia}8 simulations for unpolarized CC DIS events. The theoretical uncertainties are obtained by varying the scales renormalization scale $\mu\sim p_T^{\rm jet}$ and the jet scale $\mu_J \sim p_T^{\rm jet}R$ by a factor of 2 around their
central values and taking the envelope. We observe good agreement between the resummed TMD calculation at NLL and the \textsc{Pythia}8 results.  However, the tail of the $q_T/p_T^\nu$ distribution falls slower at high $q_T/p_T^\nu$ for the \textsc{Pythia8} simulations compared to the resummed TMD result. This is likely due to multi-jet events, which are not included as a matching contribution in the TMD result at large $q_T$.  

Figure~\ref{fig:zh_and_jt} shows our theoretical results including QCD scale uncertainties (we again take the envelope of the results when varying the scales $\mu\sim p_T^{\rm jet}$ and $\mu_J\sim p_T^{\rm jet}R$ by a factor of 2 around their central values) for the longitudinal $z_h$ and transverse momentum $j_T$ distributions for $\pi^\pm$ compared to the \textsc{Pythia8} results. We use the same simulated event sample as described above, and we observe reasonable agreement between the two results.

\begin{figure}
    \centering
    \includegraphics[width=0.94\columnwidth]{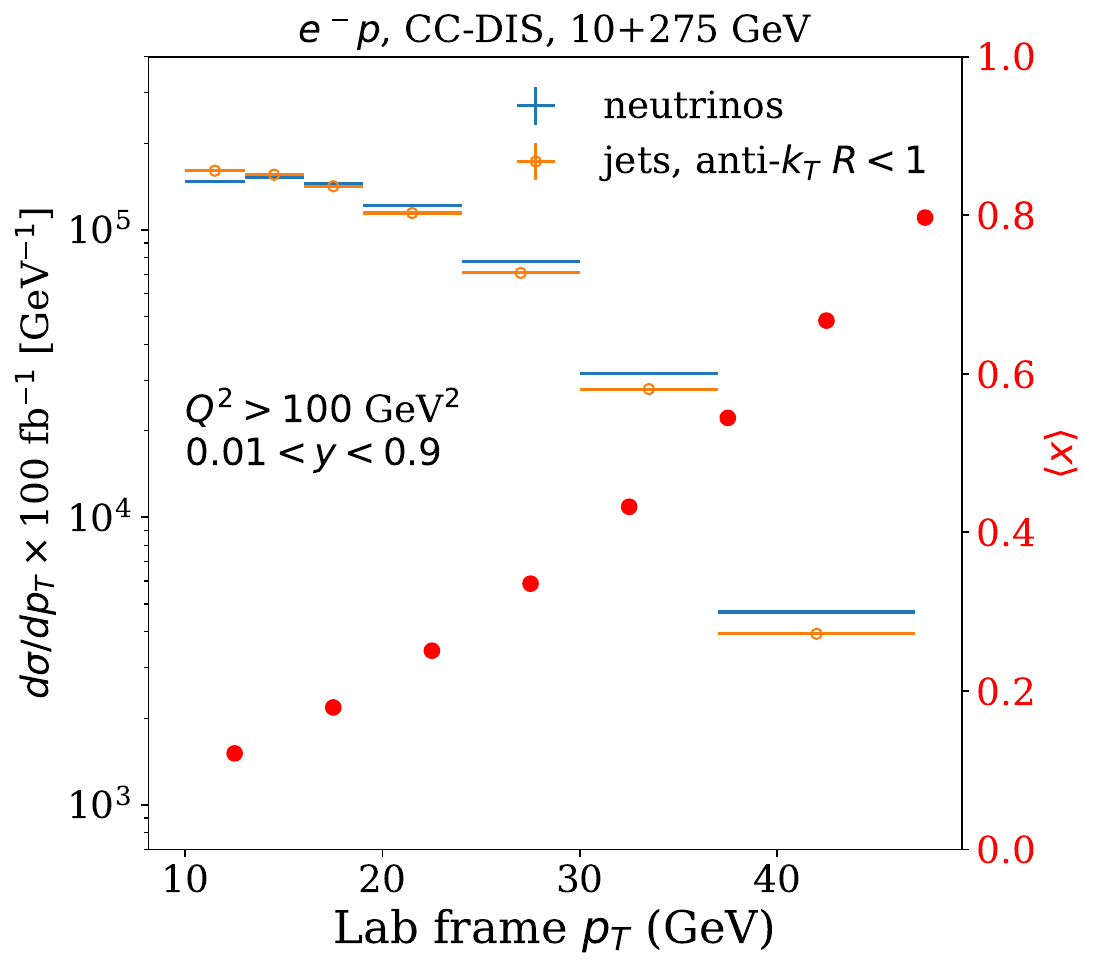}
\caption{Expected yield of neutrinos and jets in CC DIS events with an electron beam and 100 fb$^{-1}$ integrated luminosity. In addition, we show the average parton momentum fraction $x$, which is probed as a function of the neutrino transverse momentum in the laboratory frame. The cross sections generated in \textsc{Pythia}8 have been scaled to match the total cross section calculated at NLO in Ref.~\cite{Aschenauer:2013iia}.
\label{fig:cross_section}
}
\end{figure}

Lastly, Figure~\ref{fig:cross_section} shows the neutrino yields expected for 100 fb$^{-1}$, which can be collected in about a year of running at 10$^{34}$~cm$^{-2}$s$^{-1}$, as a function of the neutrino's transverse momentum.  We also show the mean of the parton momentum fraction $x$ as a function of transverse momentum (red dots). Values up to $x=0.8$ can be probed with jet/neutrino transverse momenta of $p_T=45$~GeV, which corresponds to the kinematic limit. With 100 fb$^{-1}$, the statistical uncertainty of the cross-section measurement is expected to be negligible over the entire kinematic range. However, high luminosity is needed to measure the corresponding spin asymmetries, as will be further discussed in Sec.~\ref{sec:stat} below.

\subsection{Detector-response simulations}
\label{sec:detector}

\begin{figure}
    \centering
   \includegraphics[width=0.42\textwidth]{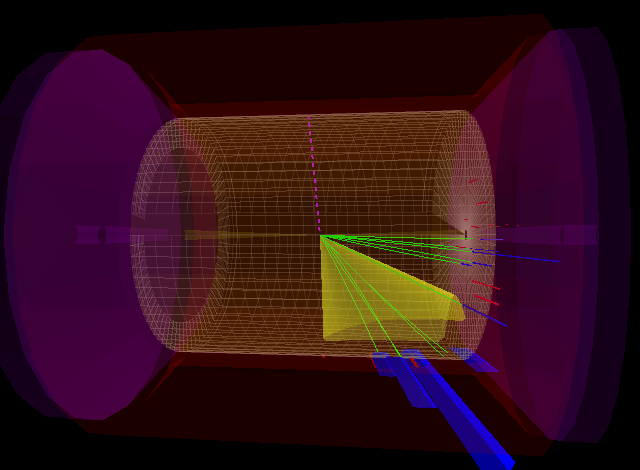}
      \includegraphics[width=0.42\textwidth]{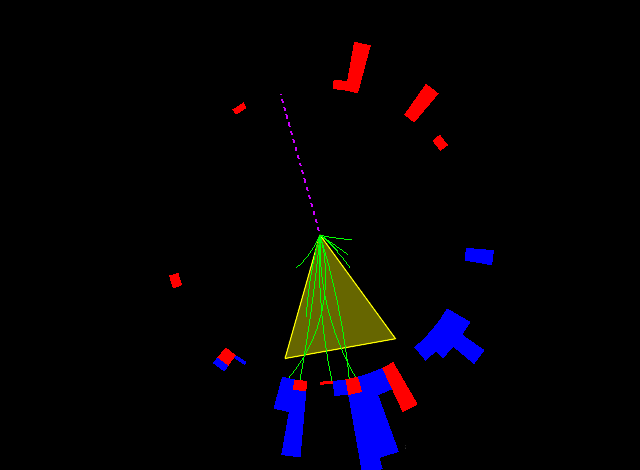}
   \caption{Display of a simulated CC DIS event using \textsc{Delphes}~\cite{deFavereau:2013fsa}.  Top:  3D view.  Bottom: Transverse view.}
   \label{fig:eventdisplay}
\end{figure}

We use the \textsc{Delphes} package~\cite{deFavereau:2013fsa} to perform fast detector simulations with parameters specified in Ref.~\cite{Arratia:2021uqr}. We consider a general-purpose detector geometry including tracking, electromagnetic and hadronic calorimeters with coverage up to $|\eta|=4.0$ and full azimuthal coverage, as described in the EIC Yellow Report~\cite{AbdulKhalek:2021gbh}. This is in line with proposed EIC detector design~\cite{Adkins:2022jfp,ATHENA:2022hxb,CORE:2022rso} that considered a high degree of hermeticity, which can be ensured with dedicated detectors at forward angles~\cite{Arratia:2022quz}. We show a representative charged-current event in Fig.~\ref{fig:eventdisplay}. 

To reconstruct jets in the detector-response simulation, we use again the \textsc{Fastjet}3.3 package~\cite{Cacciari:2011ma} with the anti-$k_T$ algorithm~\cite{Cacciari:2008gp} and $R=1$~\cite{Newman:2013ada}. As input to the jet algorithm, we use the set of particle-flow objects reconstructed with \textsc{Delphes}.

\begin{figure}
    \centering
    \includegraphics[width=0.87\columnwidth]{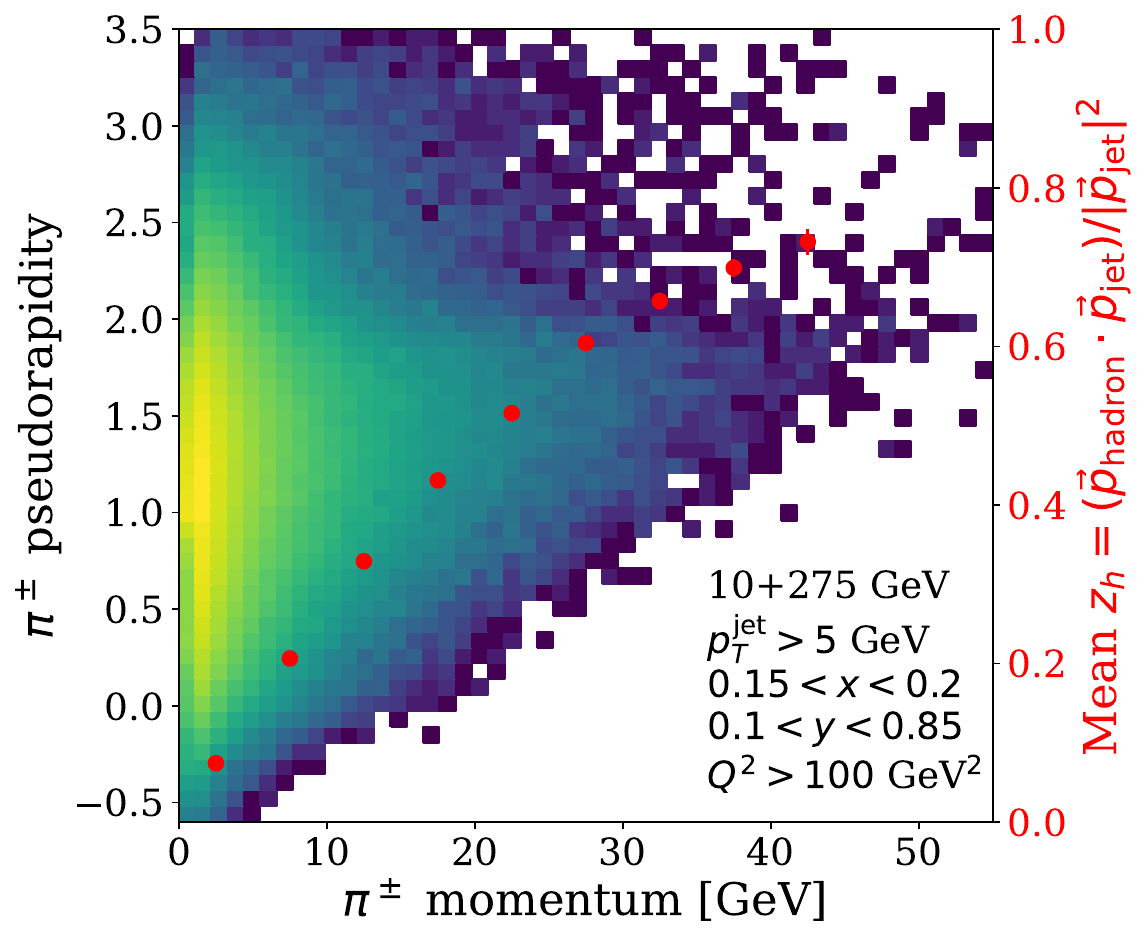}
\caption{Pseudorapidity and momentum distribution of charged pions in jets with $p_T>5$~GeV in $ep$ CC DIS. The average longitudinal-momentum fraction of the hadron with respect to the jet axis is shown by the red dots.}
\label{fig:hadron_momentum_eta_zh}
\end{figure}

In Fig.~\ref{fig:hadron_momentum_eta_zh}, we show the hadron-in-jet momenta for reconstructed $\pi^\pm$, as well as the average $z_h$ in each momentum bin.  We find that the charged pions in jets are mostly in $-$0.5$<\eta<$3.5, and have momenta up to about 45 GeV, which can be identified with high purity using gas-based Cherenkov detectors~\cite{AbdulKhalek:2021gbh}.  

\section{Event reconstruction and kinematic resolution}
\label{sec:resolution}

As typically done at particle colliders, neutrinos can be identified by measuring the missing transverse momentum, $\vec p_T^{\rm \, miss}$, which is defined as the vector sum of the transverse momenta of all measured  particles (identified using the particle-flow algorithm to avoid double-counting). 

In this section we estimate the performance of this reconstruction method for EIC. We expect this estimate to be reasonable given that the  \textsc{Delphes} fast smearing  was shown to reproduce reasonably well the performance obtained from a comprehensive detector simulation of the CMS experiment down to about $|\pTmiss|=20$ GeV~\cite{deFavereau:2013fsa}. 

\begin{figure}
    \centering
    \includegraphics[width=\columnwidth]{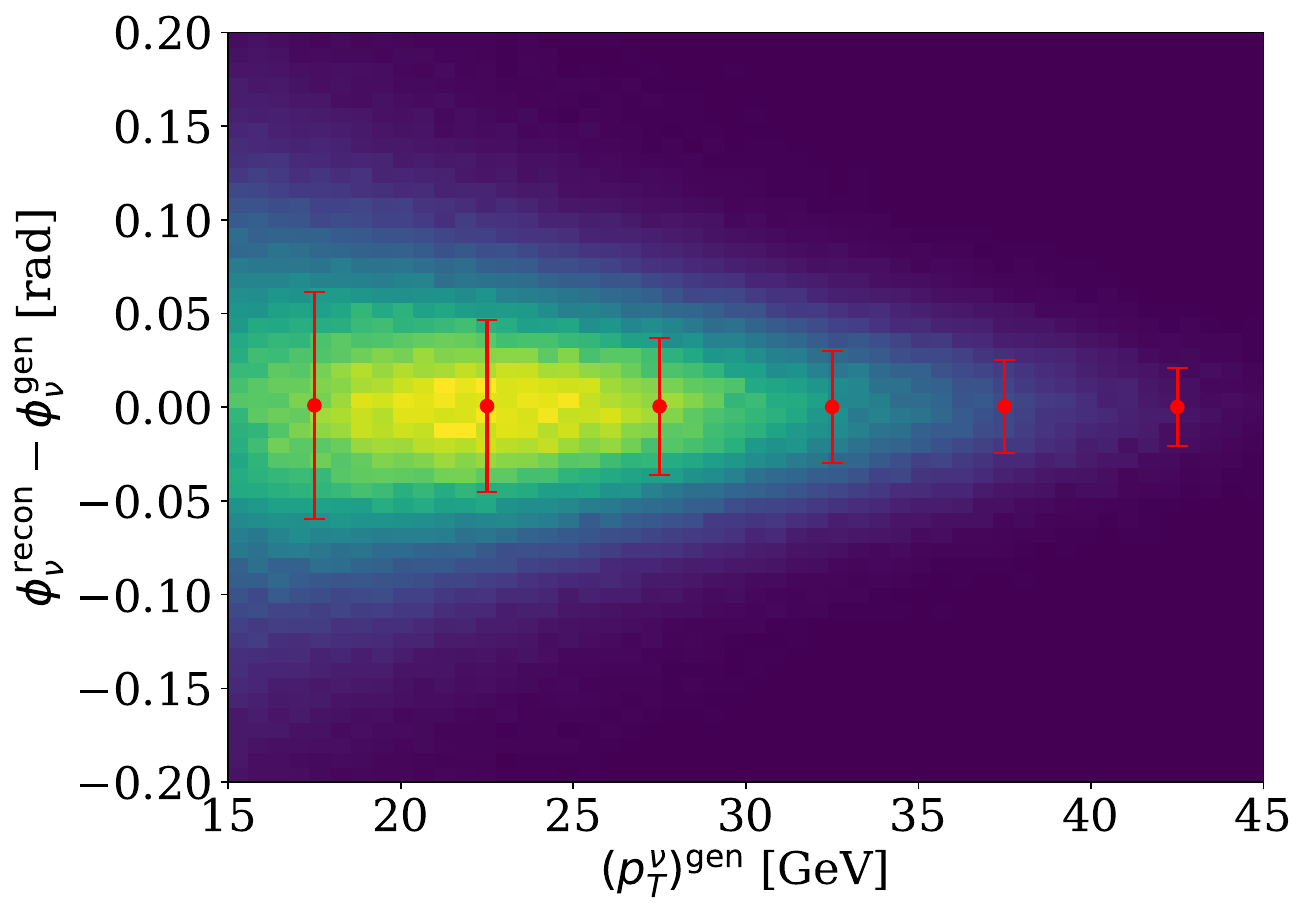}
\caption{Performance of the reconstruction of $\phi_\nu$ in CC DIS events. The red error bars indicate the means and standard deviations for given slices of $p_T^\nu$~\label{fig:phi_nu_res}}
\end{figure}

We define $\phi_\nu$ as the azimuthal angle of $-\vec p_T^{\rm \, miss}$.  We show the reconstruction performance of $\phi_\nu$ in Fig.~\ref{fig:phi_nu_res}.  The standard deviation is less than 0.06 radians, which is of similar order as the di-jet azimuthal-angle resolution of the measurement presented in Ref.~\cite{Abelev:2007ii}.

We employ the  Jacquet-Blondel (JB) method of Ref.~\cite{Amaldi:1979yh} to reconstruct the lepton kinematics. The event inelasticity is given by $y_{JB}=\sum (E_i -p_{z,i})/(2E_e)$, where the sum is over all the reconstructed particles. The four-momentum transfer is given by $Q^2_{JB}=(p^{\rm miss}_T)^2/(1-y_{JB})$ and the Bjorken scaling variable is $x_{JB} =Q^2_{JB}/(sy_{JB})$, where $s=4E_eE_p$ and $E_e$ ($E_p$) is the energy of the electron (proton) beam. The resolution of reconstructing these variables for inclusive DIS was investigated in Ref.~\cite{Aschenauer:2013iia}, and was found to be reasonable for all three of these variables. The performance of the Jacquet-Blondel method might be improved with Machine-Learning methods such as those proposed in Refs.~\cite{Arratia:2021tsq,Diefenthaler:2021rdj,Arratia:2022wny}. 

\begin{figure}
    \centering
    \includegraphics[width=0.9\columnwidth]{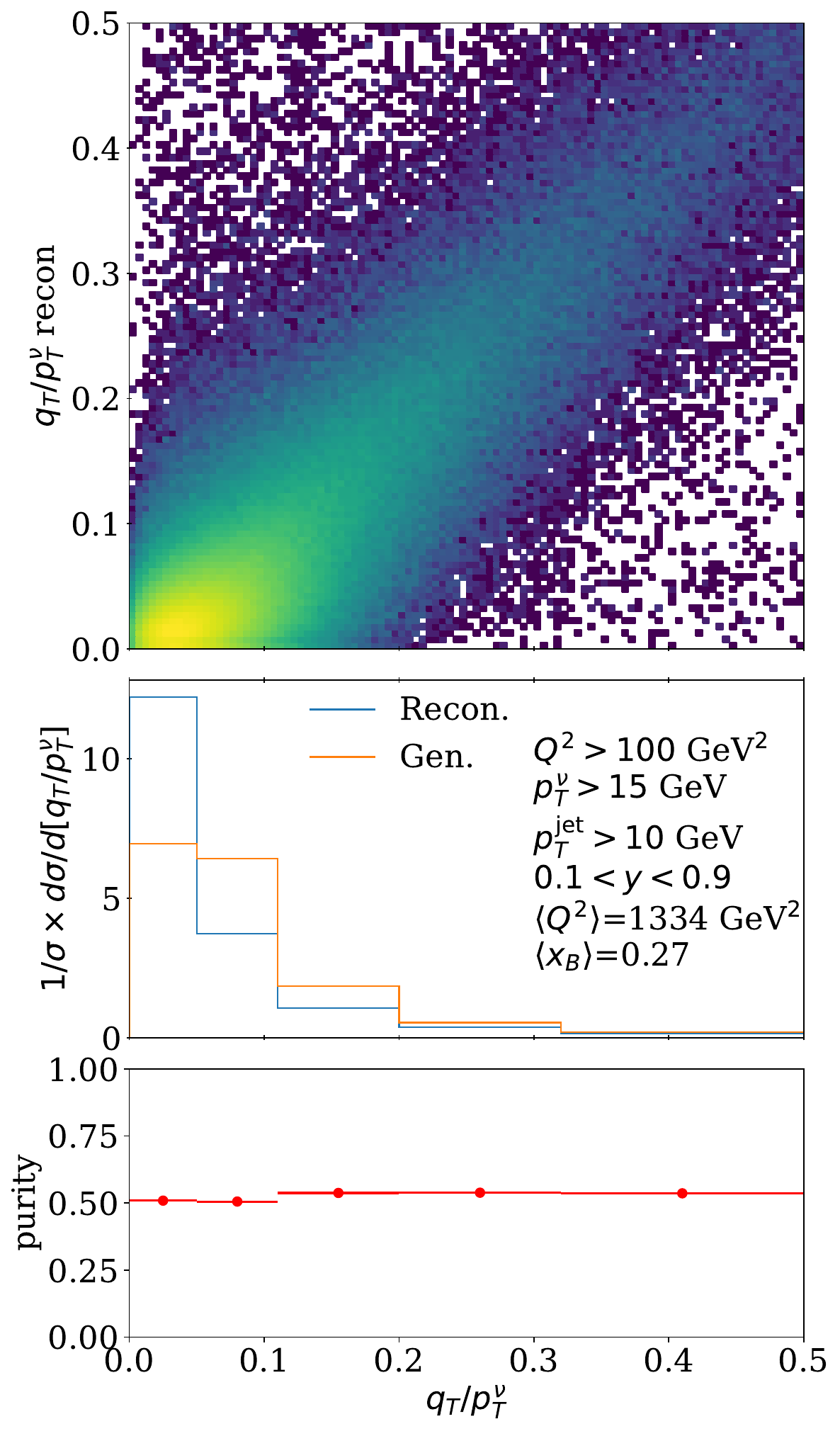}
\caption{Top: 2D histogram of the generated $q_T/p_T^\nu$ ($x$ axis) vs.~the reconstructed value ($y$ axis). Middle: Spectra of reconstructed and generated $q_T/p_T^\nu$. Bottom: Purity as a function of $q_T/p^\nu_T$.~\label{fig:qtoerptnu}}
\end{figure}

In Fig.~\ref{fig:qtoerptnu}, we compare the reconstructed values of $q_T/p_T^\nu$ with the value obtained at generator level.  In the bottom panel of this figure, we show that the ``bin purity'', or the fraction of events generated in a given bin that are reconstructed to be in the same bin. The purity is more than 50\%, which is a level amenable to standard unfolding methods. 

 \section{Suppression of the background from NC DIS and photoproduction}
 \label{sec:backgrounds}
 
 Given the relatively low rate of charged-current DIS events relative to neutral-current DIS and photo-production, the background suppression generally represents a significant challenge. If the scattered electron is missed, the event topologies of neutral- and charged-current DIS can become identical. We expect that this scenario will be significantly suppressed at the EIC compared to the HERA experiments due to improved low-angle taggers for low-$Q^{2}$ events~\cite{AbdulKhalek:2021gbh}, although the performance of these systems is hard to estimate at this point.
 
 Rather than using a low-angle scattering veto to suppress photoproduction, we follow the approach used by the CC DIS analyses at HERA~\cite{Collaboration:2010xc} which relied on two kinematic variables: $\delta=\sum_i E_i -p_{z,i}$ (where $E_i$ and $p_{z,i}$ are the reconstructed energy and longitudinal momentum of detected particles, and the sum runs over all reconstructed particles) and the ratio of the anti-parallel component $V_{AP}$ and the  parallel component, $V_P$, of the hadronic final state. The two components are defined as
\begin{equation}
V_{AP} =-\sum_i \vec p_{T,i}\cdot\hat n,\;\;\;\;{\rm for }\;\;\vec p_{T,i}\cdot\hat n < 0 \,,
\end{equation}
and 
\begin{equation}
V_{P} = \sum_i \vec p_{T,i}\cdot\hat n,\;\;\;\;{\rm for }\;\;\vec p_{T,i}\cdot\hat n > 0 \,.
\end{equation}
Here $\vec p_{T,i}$ are the transverse parts of the individual particles' momenta, $\hat n = -\vec p^{\,\nu}_T/|\vec p^{\,\nu}_T|$, and we sum over all reconstructed particles in the event. The purpose of the cuts on this variable is to ensure an azimuthally collimated energy flow. For charged-current events, the ratio $V_{AP}/V_P$ is small -- in particular for the events that we are interested in for TMD studies.  

In order to test the efficacy  of these variables for background reduction, we ran simulations of photoproduction reactions in the same manner as our CC DIS simulations, see Sec.~\ref{sec:simulation} above. We focus on photoproduction because it is expected to be the dominant background, based on experience from HERA~\cite{Collaboration:2010xc}. 

We used the following cuts: $p^{\,\nu}_T>15$ GeV, $V_{AP}/V_P<0.35$, and $\delta<30$ GeV, which are similar to the values used in Ref.~\cite{Collaboration:2010xc}.  We found that $\approx30\%$ of the generated CC~DIS events passed the cuts, whereas only 0.0005$\pm$0.0002\% of photoproduction events passed the cuts.  However, the photoproduction cross section is three orders of magnitude larger compared to CC~DIS (58~nb, compared to 14~pb, estimated using \textsc{Pythia8}).  Therefore, we estimate that about $8\pm 3\%$ of the identified event sample would be background from photoproduction when using only cuts on kinematic variables and no additional low-angle electron tagger.

Given that our estimate suggests that the background will be reduced to manageable levels, we neglect it from the projections we show in Sec.~\ref{sec:stat}.

\section{Statistical precision of spin asymmetry measurements}
\label{sec:stat}

\begin{figure}
    \centering
    \includegraphics[width=0.8\columnwidth]{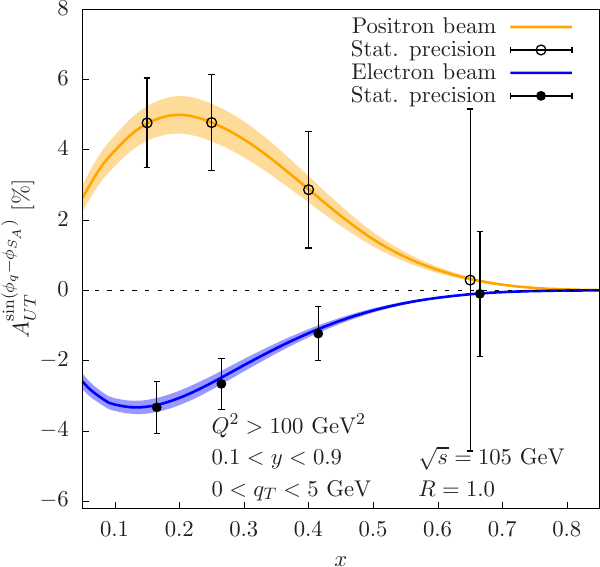}
\caption{Projected statistical precision for the neutrino-jet asymmetry, which is sensitive to the Sivers distribution, for $e^+p$ collisions (open circles) and $e^-p$ collision (closed circles), for 100 fb$^{-1}$.  The yellow and blue curves show theoretical results and the corresponding bands show the uncertainty of the extracted Sivers
function in Ref.~\cite{Echevarria:2020hpy}.}
\label{fig:sivers_theory}
\end{figure}

In Fig.~\ref{fig:sivers_theory}, we show the statistical uncertainty projected for the transverse single-spin asymmetry  $A^{\sin(\phi_q-\phi_{S_A})}_{UT}$ as a function of Bjorken $x$. Here we assume a luminosity of 100 fb$^{-1}$ and the absolute uncertainty of the asymmetry measurement is estimated to be $\sqrt{2}/(p\sqrt{N})$, where $p$ is the polarization of the proton beam, which we take to be 70$\%$, and $N$ is the number of events in a given bin that pass our cuts, scaled to match the NLO total inclusive cross section of Ref.~\cite{Aschenauer:2013iia} and an integrated luminosity of 100 fb$^{-1}$. Following Ref.~\cite{Anselmino:2011ay}, we include a factor of $\sqrt{2}$ to account for the fitting of the azimuthal modulations.  

We compare these results to the numerical results of our calculations, see Sec.~\ref{sec:theory}, which are integrated over the transverse-momentum imbalance $0 < q_T < 5$ GeV and inelasticity $0.1 < y < 0.9$. The uncertainty bands of the calculations show the uncertainty of current extractions of the Sivers function, see Ref.~\cite{Echevarria:2020hpy}. The projected statistical error bars are smaller than the predicted asymmetry for the first three bins, allowing the proposed measurement to provide a decent comparison to theoretical calculations.  

\begin{figure*}
    \centering
    \includegraphics[width=.8\textwidth]{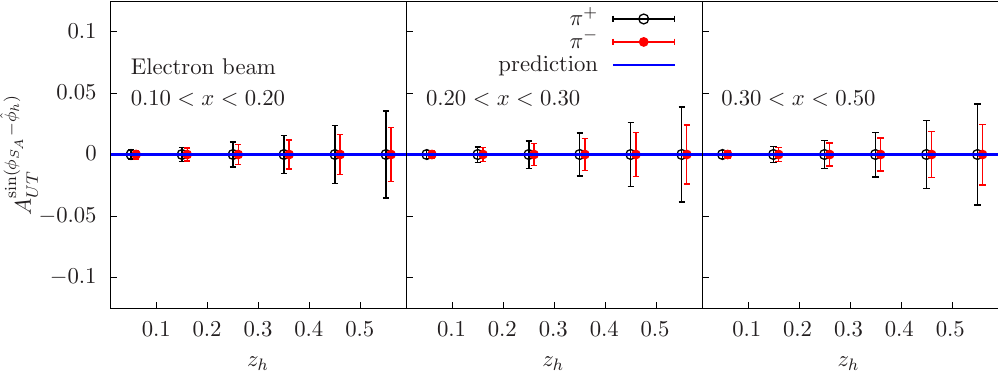}
\caption{Projected statistical precision for the $z_h$ dependence of the $\pi^\pm$-in-jet Collins asymmetries (open circles for $\pi^+$, closed circles for $\pi^-$) in electron-proton collisions.}  \label{fig:collins_theory}
\end{figure*}

While these measurements would provide weaker constraints on TMDs than analogous ones in the neutral-current channel, they offer an independent check with different flavor sensitivity. Moreover, they could test the consistency and universality of the theoretical predictions.  

The measurement that is projected in Fig.~\ref{fig:sivers_theory} would require electron and positron beams. The capability to operate with positron beam is not included in the EIC baseline design although it might be a possible upgrade. Figure~\ref{fig:sivers_theory} shows that the positron data would yield an opposite asymmetry compared to the electron data, and their comparison could help constrain in particular the TMDs associated with the $d$ quark. 

We likewise show in Fig.~\ref{fig:collins_theory} the projected statistical uncertainties for hadron-in-jet Collins asymmetries as a function of $z_h$ and Bjorken $x$ for charged pions with $j_T<1.5$~GeV. As explained in Sec.~\ref{sec:theory}, this asymmetry is expected to vanish in CC DIS within the TMD factorization formalism at leading power due to the chiral-odd nature of the transversity and Collins functions. The projected statistical uncertainties are at the level of $1\%$ or smaller for small $z_{h}$. As we have mentioned earlier, performing these measurements will enable precise tests of the theory and the assumptions of factorization and chirality of the functions involved. If the asymmetry is not observed to be exactly zero, it might indicate sensitivity to sub-leading contributions that we have neglected in Sec.~\ref{sec:theory} or some other non-standard effect.

\section{Summary and conclusions} 
\label{sec:Conclusions}
We have proposed a novel channel to study the 3D structure of the nucleon at the EIC that offers unique sensitivity to different quark flavors: neutrino-jet correlations in charge-current deep-inelastic scattering.

We have presented first calculations of unpolarized cross-sections and transverse-spin asymmetries for this channel. In addition, we performed calculations of longitudinal and transverse momentum distributions of identified hadrons inside jets and we compared our results to Monte-Carlo event-generator simulations. We find that the hadron-in-jet Collins asymmetry is exactly zero as a consequence of the chiral-odd nature of the effect. 

We used the expected EIC machine parameters in terms of luminosity and energy to estimate the kinematic reach of the proposed measurement. We also used fast detector simulations to estimate the performance of the neutrino (missing-momentum) reconstruction, and the neutrino-jet momentum imbalance. We found that these measurements should be feasible with a general-purpose detector at the EIC running at nominal luminosity.

Jet-based TMD measurements in charged-current DIS will provide important cross-checks and complement analogous measurements in the neutral-current channel. As such, we conclude that this channel represents an interesting addition to the growing science program that can be carried out with jet measurements at the future EIC.

\begin{acknowledgments}
This work was supported by the MRPI program of the University of California Office of the President, award number 00010100. M.A. was supported through DOE Contract No.~DE-AC05-06OR23177 under which JSA operates the Thomas Jefferson National Accelerator Facility. Z.K. and F.Z. are supported by the National Science Foundation under grant No.~PHY-1945471. F.R. was supported by the Simons Foundation under the Simons Bridge program for Postdoctoral Fellowships at SCGP and YITP award number 815892; the NSF, award number 1915093; the DOE Contract No.~DE-AC05-06OR23177, under which Jefferson Science Associates, LLC operates Jefferson Lab; and Old Dominion University. A.P. is supported by the National Science Foundation Grant No.~PHY-2012002 and by the DOE Contract No.~DE-AC05-06OR23177, under which Jefferson Science Associates, LLC operates Jefferson Lab.
\end{acknowledgments}

\clearpage
\appendix
\newpage

\section{Inclusive jet production}\label{app:jet}
In this appendix, we provide the expressions for all the structure functions that appear in Eq.~\eqref{eq:crosssection1}. We start with the squared matrix element of the process $e+p\rightarrow \nu+\mathrm{jet}+X$, where the incoming proton and electron are unpolarized, which is given by
\bea
&|\mathcal{M}|^2=\left(\frac{e^2}{{2}\sin^2\theta_w}\right)^2|V_{ud}|^2\frac{(g^{\mu\nu}-\frac{q^\mu q^\nu}{m_W^2})(g^{\mu'\nu'}-\frac{q^{\mu'} q^{\nu'}}{m_W^2})}{(q^2-m_W^2)^2+(m_W\Gamma_W)^2}\nnu
&\times\mathrm{Tr}\left[{\slashed{P}_D}\gamma^\mu\left(\frac{1-\gamma_5}{2}\right)\slashed{P}_B\gamma^{\mu'}\left(\frac{1-\gamma_5}{2}\right)\right]\nnu
&\times\mathrm{Tr}\left[\hat{\slashed{P}}_C\gamma^\nu\left(\frac{1-\gamma_5}{2}\right)\hat{\slashed{P}}_A\gamma^{\nu'}\left(\frac{1-\gamma_5}{2}\right)\right]\nnu
=&\, 8(G_Fm_W^2)^2|V_{ud}|^2\frac{1}{(\hat{t}-m_W^2)^2+(m_W\Gamma_W)^2}\nnu
&\times\mathrm{Tr}\left[\slashed{P}_D\gamma^\mu\left(\frac{1-\gamma_5}{2}\right)\slashed{P}_B\gamma^{\nu}\left(\frac{1-\gamma_5}{2}\right)\right]\nnu
&\times\mathrm{Tr}\left[\hat{\slashed{P}}_C\gamma_\mu\left(\frac{1-\gamma_5}{2}\right)\hat{\slashed{P}}_A\gamma_{\nu}\left(\frac{1-\gamma_5}{2}\right)\right]\,.
\eea
Here $\hat{{P}}_A=xP_A$ and $\hat{{P}}_C={P}_J$ and we used $4G_F/\sqrt{2}=e^2/(2 m_W^2\sin^2\theta_w)$. For a longitudinally polarized proton with helicity $\lambda_p$, we substitute $\hat{\slashed{P}}_A\rightarrow \gamma_5\hat{\slashed{P}}_A$. For a transversely polarized proton with transverse spin $S_T^i$, we have $\hat{\slashed{P}}_A\rightarrow \gamma_5\gamma_i\hat{\slashed{P}}_A$. However, note that the trace of the hadronic tensor vanishes for a transversely polarized proton.
For a longitudinally polarized electron with helicity $\lambda_e$, one substitutes $\slashed{P}_B\rightarrow\slashed{P}_B+\lambda_e\gamma_5\slashed{P}_B$.
The leptonic tensor is given in terms of the momenta of the electron and the left-handed neutrino:
\bea
L^{\mu\nu}&=\mathrm{Tr}\left[\slashed{P}_D\gamma^\mu(1+\lambda_e\gamma_5)\slashed{P}_B\gamma^\nu\left(\frac{1-\gamma_5}{2}\right)\right]\nnu
&=(1-\lambda_e)\left(P_B^\mu P_D^\nu+P_B^\nu P_D^\mu-g^{\mu\nu}P_B\cdot P_D+i\epsilon^{\mu\nu P_BP_D}\right)\nnu
&=L_u^{\mu\nu}+L_p^{\mu\nu}\, ,
\eea
where 
\bea
L_u^{\mu\nu}&=\left(P_B^\mu P_D^\nu+P_B^\nu P_D^\mu-g^{\mu\nu}P_B\cdot P_D+i\epsilon^{\mu\nu P_BP_D}\right)\, ,\\
L_p^{\mu\nu}&=-\lambda_e\left(P_B^\mu P_D^\nu+P_B^\nu P_D^\mu-g^{\mu\nu}P_B\cdot P_D+i\epsilon^{\mu\nu P_BP_D}\right),\nnu
\eea
represent the polarized and unpolarized components of the leptonic tensor. We can then obtain the differential cross section given in Eq.~\eqref{eq:crosssection1} with the following structure functions
\bea
F_{UU}=&\sum_q \frac{|\overline{\mathcal{M}}_{eq\rightarrow\nu q'}|^2}{16\pi^2\hat{s}^2}H(Q,\mu)\,\mathcal{J}_q(p_T^{\mathrm{jet}}R,\mu)\nnu
&\times\int\frac{{\rm d}b_T b_T}{2\pi}J_0(q_Tb_T)\,f_1^{\mathrm{TMD}}(x,b_T,\mu, \zeta)\nnu
& \qquad\times S_q(b_T,y_{\mathrm{jet}},R,\mu)\, ,\nnu
=&\,\mathcal{C}\left[f_1\right]_{eq\rightarrow\nu q'}\, ,\\
F_{LU}=&\,\mathcal{C}\left[f_1\right]_{e_Lq\rightarrow\nu q'}\, ,\label{eq:flu}\\
F_{UL}=&\,\mathcal{C}\left[g_{1L}\right]_{eq_L\rightarrow\nu q'}\, ,\\
F_{LL}=&\,\mathcal{C}\left[g_{1L}\right]_{e_Lq_L\rightarrow\nu q'}\, ,
\eea
\bea
F_{UT}^{\cos(\phi_{q}-\phi_{S_A})}=&\sum_q \frac{|\overline{\mathcal{M}}_{eq_L\rightarrow\nu q'}|^2}{16\pi^2\hat{s}^2}H(Q,\mu)\,\mathcal{J}_q(p_T^{\mathrm{jet}}R,\mu)\nnu
&\times\int\frac{{\rm d}b_T b_T^2}{4\pi M}J_1(q_Tb_T)\,g_{1T}^{(1),\mathrm{TMD}}(x,b_T,\mu, \zeta)\nnu
& \qquad\times S_q(b_T,y_{\mathrm{jet}},R,\mu)\, ,\nnu
=&\tilde{\mathcal{C}}\left[g_{1T}\right]_{eq_L\rightarrow\nu q'}\, ,\\
F_{LT}^{\cos(\phi_{q}-\phi_{S_A})}=&\,\tilde{\mathcal{C}}\left[g_{1T}\right]_{e_Lq_L\rightarrow\nu q'}\, ,\\
F_{UT}^{\sin(\phi_{q}-\phi_{S_A})}=&\,\tilde{\mathcal{C}}\left[f^\perp_{1T}\right]_{eq\rightarrow\nu q'}\, ,\\
F_{LT}^{\sin(\phi_{q}-\phi_{S_A})}=&\,\tilde{\mathcal{C}}\left[f^\perp_{1T}\right]_{e_Lq\rightarrow\nu q'}\, .
\eea

The relevant leading-order matrix elements squared are given by
\bea
|\overline{\mathcal{M}}_{eu\rightarrow\nu d}|^2
=&\, 8(G_F m_W^2)^2|V_{ud}|^2\frac{\hat{s}^2}{(\hat{t}-m_W^2)^2+m_W^2\Gamma_W^2}\,,\\
|\overline{\mathcal{M}}_{e\bar{d}\rightarrow\nu \bar{u}}|^2
=&\,8(G_F m_W^2)^2|V_{ud}|^2\frac{\hat{u}^2}{(\hat{t}-m_W^2)^2+m_W^2\Gamma_W^2} \,,\\
|\overline{\mathcal{M}}_{e_Lq\rightarrow\nu q'}|^2 =&\,-|\overline{\mathcal{M}}_{eq\rightarrow\nu q'}|^2\,,\\ \label{eq:eLq} 
|\overline{\mathcal{M}}_{eq_L\rightarrow\nu q'}|^2 =&\,-|\overline{\mathcal{M}}_{eq\rightarrow\nu q'}|^2\,,\\
|\overline{\mathcal{M}}_{e_Lq_L\rightarrow\nu q'}|^2 =&\,|\overline{\mathcal{M}}_{eq\rightarrow\nu q'}|^2\,, \label{eq:eLqL} 
\eea
The matrix elements for an unpolarized and polarized electron are related to each other, see Eqs.~\eqref{eq:eLq}, \eqref{eq:eLqL} due to the factor $(1-\lambda_e)$ in the expression of the leptonic tensor. As a result, we obtain the following relations between the different structure functions:
\bea
F_{LU}&=-F_{UU}\,,\\
F_{LL}&=-F_{UL}\,,\\
F_{LT}^{\cos(\phi_q-\phi_{S_A})}&=-F_{UT}^{\cos(\phi_q-\phi_{S_A})}\,,\\
F_{LT}^{\sin(\phi_q-\phi_{S_A})}&=-F_{UT}^{\sin(\phi_q-\phi_{S_A})}\,.
\eea

\section{Hadron distributions inside the jet~\label{app:h-in-jet}}

For an unpolarized final-state hadron, we obtain TMD JFFs $\mathcal{D}_1,\ \mathcal{H}_1^\perp$ at leading-twist~\cite{Kang:2021ffh}. The corresponding correlator can be written as follows
\bea\label{eq:jT_direction_dependent}
\Delta(z_h,\vec{j}_T)=&\mathcal{D}_1^{h/q}(z_h,{j}_T)\frac{\slashed{n}_-}{2}\nnu
&-i\mathcal{H}^{\perp,h/q}_1(z_h,{j}_T)\frac{\slashed{j_T}}{z_hM_h}\frac{\slashed{n}_-}{2}\, ,
\eea
where we suppress the depenence on the renormalization scale $\mu$ and the Collins-Soper scale $\zeta$~\cite{Collins:2011zzd}. The different traces of the correlator are given by
\bea
\Delta^{h/q[\gamma^-]}=&\mathcal{D}_1^{h/q}(z_h,{j}_T)\, ,\\
\Delta^{h/q[i\sigma^{i-}\gamma_5]}=&\frac{\epsilon_T^{ij}{j_T}^j}{z_hM_h}\mathcal{H}^{\perp,h/q}_1(z_h,{j}_T)\, ,
\eea
For an electron colliding with an unpolarized or a longitudinally polarized initial proton, we obtain the same partonic scattering amplitudes as shown in Appendix~\ref{app:jet}.  However, if the electron collides with a transversely polarized quark from the initial proton, the corresponding term in the hadronic tensor is given by 
\bea
H^{\mu\nu}_{\mathcal{H}_1^{\perp h/q}}={\rm Tr}\left[\frac{\slashed{j_T}}{z_hM_h}\hat{\slashed{P}}_C\gamma^\mu\left(\frac{1-\gamma_5}{2}\right)\slashed{v}\hat{\slashed{P}}_A\gamma^\nu\left(\frac{1-\gamma_5}{2}\right)\right]\,
\label{eq:had_tensor}
\eea
where $\hat{{P}}_A=xP_A$ and $\hat{{P}}_C={P}_J$. For different TMD PDFs, the vector $\slashed{v}$ is given by $\big(-\slashed{S}_T\gamma_5\big)$ for $h_1$, $\big(-\lambda_p\slashed{k}_T\gamma_5/M\big)$ for $h_{1L}^\perp$, $\big(-i\slashed{k}_T/M\big)$ for $h_1^\perp$ and $\big((\vec{k}_T\cdot\vec{S}_T\slashed{k}_T-\vec{k}_T^2\slashed{S}_T/2)\gamma_5/M^2\big)$ for $h_{1T}^\perp$. Note that there are always three $\gamma$ matrices between the $(1-\gamma_5)/2$ factors in the expression of the hadronic tensor in Eq.~\eqref{eq:had_tensor}. Thus, we find
\bea\label{eq:g5mat}
&\left(\frac{1-\gamma_5}{2}\right)\gamma_\alpha\gamma_\beta\gamma_\rho\left(\frac{1-\gamma_5}{2}\right)\nnu
&=\gamma_\alpha\gamma_\beta\gamma_\rho\left(\frac{1+\gamma_5}{2}\right)\left(\frac{1-\gamma_5}{2}\right)=0\,.
\eea
Therefore, the expression in Eq.~\eqref{eq:had_tensor} vanishes and spin asymmetries involving transversely polarized quarks in CC DIS are zero. As a result, all contributions related to chiral-odd Collins jet fragmentation function do not appear in the differential cross section for hadron-in-jet production. Here, we show the differential cross section in terms of the remaining non-zero structure functions
\bea
&\frac{{\rm d}\sigma^{ep\rightarrow \nu+\mathrm{jet}X}}{{\rm d}y_J\, {\rm d}^2p_{JT}\, {\rm d}^2q_T\,{\rm d}z_h\,{\rm d}^2{j_T}}=F^h_{UU}+\lambda_p F^h_{UL}\nnu
&+|S_T|\left[\cos(\phi_{q}-\phi_{S_A})F_{UT}^{h,\cos(\phi_{q}-\phi_{S_A})}\right.\nnu
&\hspace{1.2cm}\left.+\sin(\phi_{q} -\phi_{S_A})F_{UT}^{h,\sin(\phi_{q}-\phi_{S_A})}\right]\nnu
&+\lambda_e\left[F^h_{LU}+|S_T|\sin(\phi_{q}-\phi_{S_A})F_{LT}^{h,\sin(\phi_{q}-\phi_{S_A})}\right.\nnu
&\quad+\left.\lambda_pF^h_{LL}+|S_T|\cos(\phi_{q}-\phi_{S_A})F_{LT}^{h,\cos(\phi_{q}-\phi_{S_A})})\right]\,.
\eea
In total there are 8 structure functions, which are given by
\bea
F^h_{UU}=&\sum_q \frac{|\overline{\mathcal{M}}_{eq\rightarrow\nu q'}|^2}{16\pi^2\hat{s}^2}H(Q,\mu)\,\mathcal{D}_{1}(p_T^{\mathrm{jet}}R,\mu)\nnu
&\times\int\frac{{\rm d}b_Tb_T}{2\pi}J_0(q_Tb_T)\,f_1^{\mathrm{TMD}}(x,b_T,\mu, \zeta)\nnu
&\times S_q(b_T,y_{\mathrm{jet}},R,\mu)\, ,\nnu
=&\,\mathcal{C}^h\left[f_1\mathcal{D}_1\right]_{eq\rightarrow\nu q'}\, ,\\
F^h_{LU}=&\,\mathcal{C}^h\left[f_1\mathcal{D}_1\right]_{e_Lq\rightarrow\nu q'}\, ,\\
F^h_{UL}=&\,\mathcal{C}^h\left[g_{1L}\mathcal{D}_1\right]_{eq_L\rightarrow\nu q'}\, ,\\
F^h_{LL}=&\,\mathcal{C}^h\left[g_{1L}\mathcal{D}_1\right]_{e_Lq_L\rightarrow\nu q'}\, ,
\eea
\bea
F_{UT}^{h,\cos(\phi_{q}-\phi_{S_A})}=&\sum_q \frac{|\overline{\mathcal{M}}_{eq_L\rightarrow\nu q'}|^2}{16\pi^2\hat{s}^2}H(Q,\mu)\,\mathcal{D}_{1}(p_T^{\mathrm{jet}}R,\mu)\nnu
&\times\int\frac{{\rm d}b_T b_T^2}{4\pi M}J_1(q_Tb_T)\,g_{1T}^{(1),\mathrm{TMD}}(x,b_T,\mu, \zeta)\nnu
&\times S_q(b_T,y_{\mathrm{jet}},R,\mu)\, ,\nnu
=&\,\tilde{\mathcal{C}}^h\left[g_{1T}\mathcal{D}_1/M\right]_{eq_L\rightarrow\nu q'}\, ,\\
F_{LT}^{h,\cos(\phi_{q}-\phi_{S_A})}=&\,\tilde{\mathcal{C}}^h\left[{g_{1T}\mathcal{D}_1}/{M}\right]_{e_Lq_L\rightarrow\nu q'}\, ,\\
F_{UT}^{h,\sin(\phi_{q}-\phi_{S_A})}=&\,\tilde{\mathcal{C}}^h\left[{f^\perp_{1T}\mathcal{D}_1}/{M}\right]_{eq\rightarrow\nu q'}\, ,\\
F_{LT}^{h,\sin(\phi_{q}-\phi_{S_A})}=&\,\tilde{\mathcal{C}}^h\left[{f^\perp_{1T}\mathcal{D}_1}/{M}\right]_{e_Lq\rightarrow\nu q'}\, .
\eea

We can obtain relations between the different hadron-in-jet structure functions analogous to inclusive jets production (see Appendix~\ref{app:jet}):

\bea
F^h_{LU} &= -F^h_{UU}\,,\\
F^h_{LL}&=-F^h_{UL}\,,\\
F_{LT}^{h,\cos(\phi_q-\phi_{S_A})}&=-F_{UT}^{h,\cos(\phi_q-\phi_{S_A})}\,,\\
F_{LT}^{h,\sin(\phi_q-\phi_{S_A})}&=-F_{UT}^{h,\sin(\phi_q-\phi_{S_A})}\,.
\eea

\renewcommand\refname{Bibliography}
\bibliographystyle{utphys} 
\bibliography{bibio.bib} 

\end{document}